\documentclass[a4paper,11pt]{article}
\usepackage{pos}

\title{Two-component WIMP – FIMP dark matter}

\author*[a]{Francesco Costa}

\affiliation[a]{Institute for Theoretical Physics, Georg-August University Göttingen,\\
Friedrich-Hund-Platz 1, Göttingen D-37077, Germany}

\emailAdd{francesco.costa@theorie.physik.uni-goettingen.de}

\abstract{The document discusses a proposed extension to the Standard Model that aims to explain the presence of neutrino masses and the existence of dark matter. The model includes two potential candidates for dark matter, a vector WIMP and a fermion FIMP, and their combined presence accounts for the total amount of observed dark matter. This study examines the various ways in which dark matter could be produced within this model and explores the connections between the dark matter and neutrino sectors. It also examines various constraints from existing and future experiments. Additionally, the model includes a scalar field that can play a role in a first-order phase transition in the early universe, and the article looks at the potential for the production of gravitational waves as a result of this phase transition and their detectability. This study also assesses the possibility for this phase transition to be strong enough to drive the electroweak baryogenesis.}

\FullConference{%
  8th Symposium on Prospects in the Physics of Discrete Symmetries (DISCRETE 2022)\\
  7-11 November, 2022\\
  Baden-Baden, Germany
}

\begin{document}
\maketitle

\section{Introduction}

The Standard Model (SM) of particle physics has been successful in the past decades with experiments matching its predictions in particular with the discovery of the Higgs boson. However, it does not explain certain observations such as the non-zero mass of neutrinos~\cite{Super-Kamiokande:1998kpq} and the existence of dark matter~\cite{Planck:2018vyg}. One proposed mechanism to generate the mass of neutrinos is the type-I seesaw mechanism~\cite{Minkowski:1977sc}, which involves introducing heavy singlet leptons. An extended version of this mechanism, called the extended double seesaw~\cite{Kang:2006sn,Mitra:2011qr}, has also been proposed to achieve a low-scale leptogenesis without fine-tuning the heavy neutrino masses. This mechanism also allows for the possibility of detection at future collider experiments. 

Dark matter is another missing piece of the SM and recent studies have explored alternative mechanisms to the standard freeze-out~\cite{Gunn:1978gr} such as the freeze-in mechanism~\cite{McDonald:2001vt,Hall:2009bx} in which the dark matter particle is called a Feebly Interacting Massive Particle (FIMP) because its interaction with the SM is much smaller than the electroweak scale. The relic abundance is then produced by out-of-equilibrium scattering or decay processes. A more general multi-component DM scenario is also possible where both freeze-out and freeze-in could have been active in the early universe, contributing to the total DM relic density~\cite{Zurek:2008qg,Arcadi:2016kmk,Profumo:2009tb,Costa:2022oaa}.

\subsection{The model}

Here we explored a beyond the Standard Model (BSM) scenario that addresses the aforementioned problems~\cite{Costa:2022lpy}. The model introduces two sets of three-generation neutrinos, $N_L^i$, and $S_L^i$; the first two generations are used to explain the light neutrino masses in an extended double-seesaw mechanism, and their third generation serves as FIMP dark matter candidates. The scenario also includes a vector gauge boson $W_D$ associated with an extra dark $U(1)_D$ gauge symmetry, which also plays the role of a WIMP dark matter candidate. Additionally, the dark Higgs field $\phi_D$ associated with the extra dark $U(1)_D$ modifies the scalar sector, leading to a first-order phase transition (FOPT)~\cite{Chao:2014ina} and we discussed the detection possibilities of its associated stochastic gravitational waves (GW)~\cite{Kamionkowski:1993fg}.
The symmetries and field content are summarized in Table 1.
\begin{table}[t!]
\centering
\renewcommand{\arraystretch}{1.2}
\tabcolsep=0.11cm
\begin{tabular}{||c|c|c|c||}
\hline
\hline
\begin{tabular}{c}
Gauge\\
Group\\ 
\hline
$SU(2)_{L}$\\ 
\hline
$U(1)_{Y}$\\ 
\hline
$U(1)_{D}$\\ 
\end{tabular}
&
\begin{tabular}{c|c|c}
\multicolumn{3}{c}{Baryon Fields}\\ 
\hline
$Q_{L}^{i}=(u_{L}^{i},d_{L}^{i})^{T}$
&$u_{R}^{i}$
&$d_{R}^{i}$\\ 
\hline
$2$&$1$&$1$\\ 
\hline
$1/6$&$2/3$&$-1/3$\\ 
\hline
$0$&$0$&$0$\\ 
\end{tabular}
&
\begin{tabular}{c|c|c|c}
\multicolumn{4}{c}{Lepton Fields}\\
\hline
$L_{L}^{i}=(\nu_{L}^{i},e_{L}^{i})^{T}$ 
& $e_{R}^{i}$ 
& $N_{L}^{i}$
& $S_{L}^{i}$\\
\hline
$2$&$1$&$1$&$1$\\
\hline
$-1/2$&$-1$&$0$&$0$\\
\hline
$0$&$0$&$0$&$0$\\
\end{tabular}
&
\begin{tabular}{c|c}
\multicolumn{2}{c}{Scalar Fields}\\
\hline
$\phi_{h}$&$\phi_{D}$\\
\hline
$2$&$1$\\
\hline
$1/2$&$0$\\
\hline
$0$&$1$\\
\end{tabular}\\
\hline
\hline
\end{tabular}
\caption{Particle contents and their corresponding charges under gauge groups.}
\label{tab:contents}
\end{table}

\section{Dark matter}
 \begin{figure}
     \centering
     \includegraphics[scale=0.5]{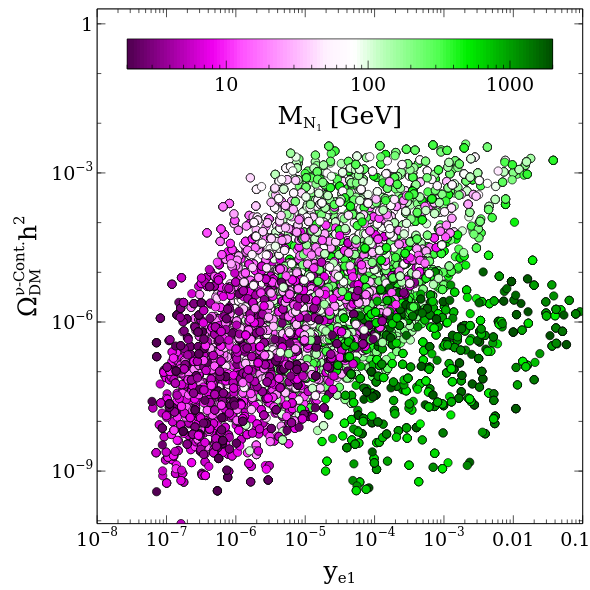}
     \includegraphics[scale=0.5]{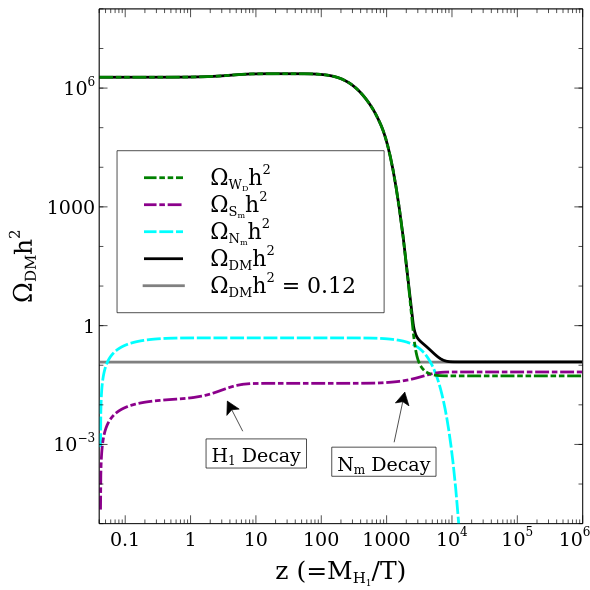}
     \caption{Left panel: Allowed parameter spaces in the $y_{e1}$ -- $\Omega^{\nu}_{\rm DM}h^{2}$ . Right panel: DM production by the freeze-out and freeze-in mechanisms and its evolution in terms of $z$. 
The model parameters are chosen as $M_{N_{m}} = 300$ GeV, $M_{S_{m}} = 20$ GeV, $M_{W_D} = 1.04628$ GeV, $\Lambda = 5.5 \times 10^{14}$ GeV, $\kappa = \kappa^{\prime} = \xi = \xi^{\prime} = \alpha = \alpha^{\prime} = 1$, $M_{H_2} = 2.212$ GeV, $g_{D} = 3.1 \times 10^{-4}$, where $H_2$ is the dark Higgs mass eigenstate and $g_D$ the dark gauge coupling. 
The green double-dot-dashed (purple dot-dashed) line corresponds to the WIMP (FIMP) DM relic density. The cyan dashed line represents the NLSP relic density. The sum of the WIMP and FIMP DM relic densities is depicted by the black solid line, while the grey solid line shows the present DM relic density measured by the Planck, $\Omega_{\rm DM}h^2 = \Omega_{\rm Tot}h^2 = 0.12$.
}
     \label{fig:neutrino-scatter-4}
 \end{figure}
We considered the productions of $S^3_{L}$ and $N^3_{L}$ to be through dimension-5 operators, which are the only ones allowed considering the two FIMP particles to be odd under a $Z_2$ symmetry. These operators get naturally suppressed when the scale of new physics $\Lambda$ is large, ensuring feeble interactions with the rest of the particle spectra; we considered $\Lambda \ge 10^{14}$ GeV.
\begin{align}
\mathcal{L}_{\rm DM} &=  \frac{\kappa}{\Lambda} S^{3}_{L} S^{3}_{L} (\phi^{\dagger}_h \phi_h)
+ \frac{\kappa^{\prime}}{\Lambda} S^{3}_{L} S^{3}_{L} (\phi^{\dagger}_D \phi_D)
+ \frac{\xi}{\Lambda} N^{3}_{L} N^{3}_{L} (\phi^{\dagger}_h \phi_h)
+ \frac{\xi^{\prime}}{\Lambda} N^{3}_{L} N^{3}_{L} (\phi^{\dagger}_D \phi_D)
\nonumber \\ 
&\quad 
+ \frac{\alpha}{\Lambda} N^{3}_{L} S^{3}_{L} (\phi^{\dagger}_h \phi_h)
+ \frac{\alpha^{\prime}}{\Lambda} N^{3}_{L} S^{3}_{L} (\phi^{\dagger}_D \phi_D)
+ {\rm h.c.}
\,.
\label{eqn:ldm}
\end{align}
We consider $S_m$ as the FIMP DM candidate, which has the lighter mass eigenvalue among the FIMP particles, $N_m$ is the other eigenstate. In a scenario where only the coupling $\kappa$ is active and the mixing between $S_L^3$ and $N_L^3$ is negligible, $S_m \sim S_L^3$. Additionally, if the mixing between the Higgses is small, meaning $\cos\theta$ is close to 1, the FIMP DM $S_m$ is mainly produced through Higgs scattering, and the analytical solution for the yield is given by~\cite{Hall:2009bx}:
\begin{align}\label{eqn:YSmapprox}
Y_{S_m}	= \int^{T_{R}}_{T_{\rm end}} \frac{1}{S\mathcal{H}T}\left(\frac{4 \kappa}{\Lambda}\right)^2 \frac{1}{16 \pi^5} \mathrm{~T}^6
\,,
\end{align}
where $T_{R}$ is the reheating temperature. To obtain the correct relic abundance and asking for $T_R > T_{EWSB}$, the electroweak symmetry breaking temperature, we found that the range of the preferred parameter is: $10^{13} < \Lambda < 10^{6}$ and $ 10^{2} <T_R< 10^{5}$. In particular, we shall choose $T_R = 3$ TeV throughout this study.
To obtain the final plot we numerically evolve the full Boltzmann equations using {\tt micrOMEGAs} to obtain the DM relic densities.

 $S_m$ can be produced also through the annihilations of active neutrinos and extra heavy neutrinos, mediated by Higgses, such as $\nu_i + N_j \xrightarrow[]{H_{1,2}} S_m + S_m$ and $\nu_i + S_j \xrightarrow[]{H_{1,2}} S_m + S_m$, where $i=1,2,3$ and $j=1,2$. These allowed channels come from the extended double seesaw neutrino sector

 \begin{equation}
\mathcal{L}_{N} \supset
- \sum_{i,j = 1,2 }\mu_{ij} S^{i}_{L} S^{j}_{L}
- \sum_{i,j = 1,2 } M_{S}^{ij} S^{i}_{L} N^{j}_{L}  
- \sum_{i,j = 1,2 } M_{R}^{ij} N^{i}_{L} N^{j}_{L}
-\sum_{i=e,\,\mu,\,\tau, j=1,2} y_{ij} \bar{L_{i}}
\tilde {\phi}_{h} N_{j} 
+ {\rm h.c.}
\label{eqn:lagN}
\end{equation}
In Figure~\ref{fig:neutrino-scatter-4}, we present the allowed parameter space for the Yukawa coupling $y_{e1}$ and the DM relic density that comes solely from the neutrino sector.  We can see that when $M_{N_1}$ is less than 500 GeV, there is a linear relationship between $y_{e1}$ and the DM relic density coming from the active and heavy neutrinos' annihilations. This reflects the fact that $\Omega^{\nu}_{\rm DM}h^2 \propto y_{e1}^2$. When $M_{N_1}$ is larger than $10^3$ GeV, the contribution to the DM relic density is small, as the mass is close to the chosen reheating temperature of $T_R = 3$ TeV, leading to suppression. We observe that for the chosen range of parameter values the contribution of the active and extra heavy neutrinos to the total DM relic density is at most about 3\%.
The results are obtained for points in the parameter space that are not excluded by lepton flavor violation bounds and are in agreement with the neutrino oscillation data.

The WIMP candidate $W_D$ is produced via the standard freeze-out mechanism and its mass should be close to the resonance to avoid overproduction. The right panel of  Fig. 1 shows the production of dark matter (DM) through freeze-out and freeze-in mechanisms. The green double-dot-dashed line represents the WIMP DM produced by freeze-out, which occurs at $T \simeq M_{W_D}/20$ or $z \simeq 2500$. The cyan dashed line represents the production of the next-to-lightest-stable-particle (NLSP) $N_m$, which later decays to the FIMP DM $S_m$ at $z \simeq 3500$. The NLSP is produced in the early Universe at $T \simeq 3000$ GeV through $2 \rightarrow 2$ processes. The purple dot-dashed line indicates the FIMP DM production via the freeze-in mechanism, with initial production at $z = 0.03$ and additional production from the decay of the SM-like Higgs, $H_1$, and the NLSP. The total DM relic density, shown by the black solid line, matches the Planck measurement of $\Omega_{\rm Tot}h^{2} = 0.12$ today, with the WIMP and FIMP DM contributing equally.

\subsection{Experimental bounds}
\begin{figure}[t!]
\centering
\includegraphics[scale=0.49]{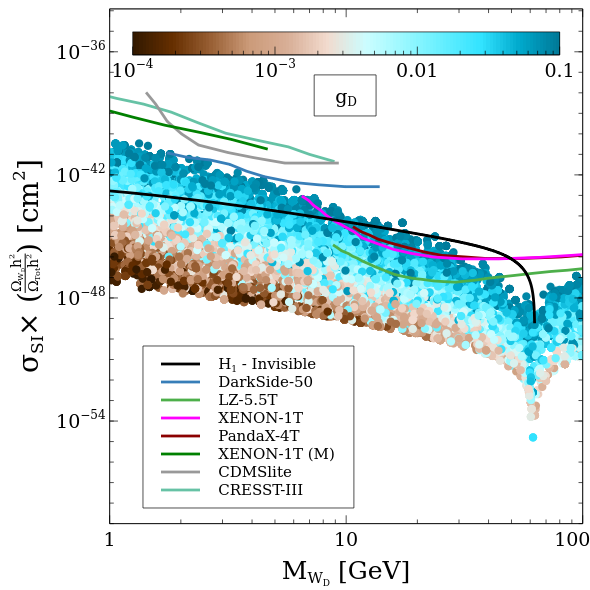}
\includegraphics[scale=0.49]{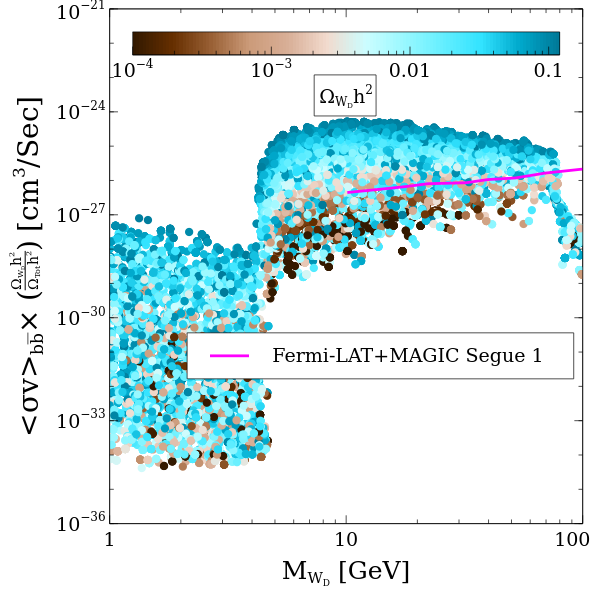}
\caption{
Allowed parameter space satisfying $0.01 \leq \Omega_{\rm DM}h^2 \leq 0.12$ in the $M_{W_D}$ -- $(\Omega_{W_D}/\Omega_{\rm Tot})\sigma_{\rm SI}$ (left) and $M_{W_D}$ -- $(\Omega_{W_D}/\Omega_{\rm Tot})\langle\sigma v\rangle_{b\bar{b}}$ (right) planes. Here, $\Omega_{\rm Tot}h^2 = 0.12$ is total DM relic density today. The black solid line in the left panel indicates the Higgs invisible decay constraint. Various direct and indirect detection bounds are also overlaid with coloured solid lines; see text for detailed explanation. The colour of the points represents the value of the dark gauge coupling $g_D$ (left) and the WIMP DM relic density (right).
} 
\label{fig:DM-scatter-plot-3}
\end{figure}

In the next section, we see that a low-mass BSM dark Higgs is favored by FOPT. Therefore, in this section, we will focus on the range of $1-200$ GeV for the dark Higgs. Additionally, to avoid potential issues with collider searches due to the low mass, we considered small mixing angles of $|\sin\theta| < 0.1$ to evade Higgs signal strength bounds.
We will consider five main constraints for the discussion of DM phenomenology: 1) relic density, 2) direct detection bounds, 3) indirect detection bounds, 4) Higgs invisible decay, and 5) Higgs signal strength bound. 

The right panel of Figure 2 shows the allowed region in the $M_{W_D}$ --  $(\Omega_{W_D}/\Omega_{\rm Tot})\sigma_{\rm SI}$ (LP) and $M_{W_D}$ --  $(\Omega_{W_D}/\Omega_{\rm Tot})\langle\sigma v\rangle_{b\bar{b}}$ planes, together with various direct and indirect detection bounds that are depicted by solid lines. Note that we have rescaled the $y$-axes by the amount of the WIMP DM relic density compared to the total DM in the Universe $\Omega_{\rm Tot} h^{2} = 0.12$. A part of the $M_{W_D} > 7$ GeV region is already ruled out by the different direct detection experiments such as XENON-1T~\cite{XENON:2018voc}. 
The region of DM mass below 7 GeV will be explored by future experiments like DarkSide-50~\cite{DarkSide:2018kuk}. 
 The region above the black solid line is already ruled out by the current bound on the branching of the Higgs invisible decay mode.
 The region of $M_{W_D}\gtrsim 10$ GeV is constrained by the Fermi-LAT + MAGIC Segue 1 data~\cite{MAGIC:2016xys}. We observe that part of the parameter space which contributes dominantly to the DM relic is already ruled out by the indirect detection bound. Future experiments will be able to further test the allowed parameter space.

\section{First order phase transition}
The extra dark $U(1)_D$ Higgs field not only gives a mass to the WIMP DM $W_D$, but it also changes the vacuum evolution. To study the potential we consider only the temperature corrections and we neglect the Coleman Weinberg terms that would introduce renormalization scale and gauge dependence~\cite{Chiang:2018gsn,Nielsen:1975fs}.

Considering the VEV of the $U(1)_D$ Higgs to be non-zero at zero temperature we have two options for the phase transition pattern: the one-step phase transition has the pattern $(\langle H \rangle, \langle H_D \rangle) = (0,0) \rightarrow (v,v_D)$, while the two-step phase transition may occur via $(\langle H \rangle, \langle H_D \rangle) = (0,0) \rightarrow (0,v_D^\prime) \rightarrow (v,v_D)$ or $(\langle H \rangle, \langle H_D \rangle) = (0,0) \rightarrow (v^\prime,0) \rightarrow (v,v_D)$.
For the two-step phase transition of the pattern $(\langle H \rangle, \langle H_D \rangle) = (0,0) \rightarrow (0,v_D^\prime) \rightarrow (v,v_D)$, the second step breaks the electroweak symmetry, giving~\cite{Carena:2019une}
\begin{align}\label{eqn:vcTcExpr}
\frac{v_c}{T_c} = 
\frac{2E^{\rm SM}}{\lambda_h - \lambda_{hD}^2/(4\lambda_D)}
= \frac{4 E^{\rm SM} v^2}{M_{H_1}^2}\left(1+\sin ^2 \theta\, \frac{M_{H_1}^2-M_{H_2}^2}{M_{H_2}^2}\right)\,.
\end{align}
Strong FOPTs, $v_c/T_c\gtrsim 1$, are then achieved for small values of $\lambda_m$, or equivalently, small values of the dark $U(1)_D$ Higgs mass. Therefore if we consider this part of the parameter space the model satisfies one of the necessary conditions for successful electroweak baryogenesis.

\begin{figure}
    \centering
    \includegraphics[scale=0.6]{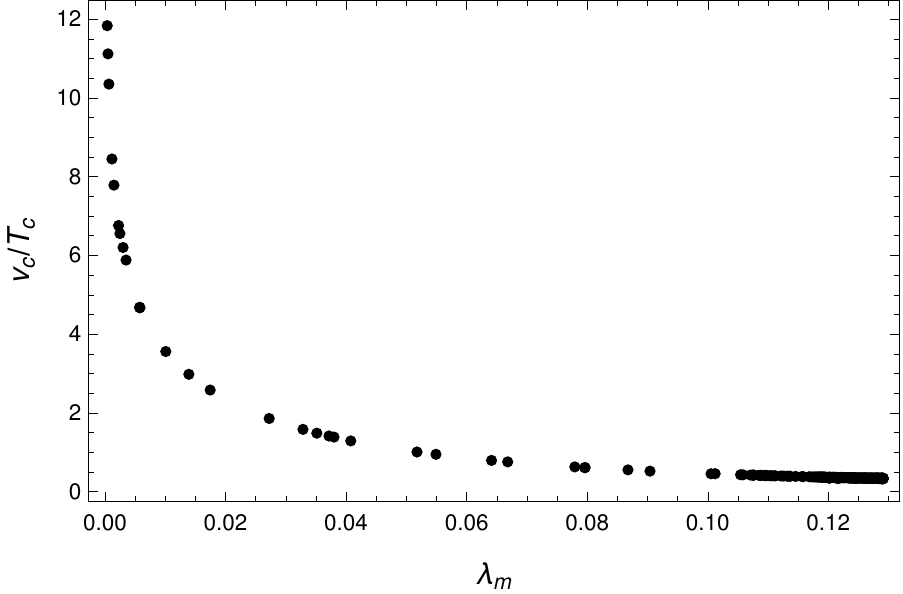}
    \includegraphics[scale=0.6]{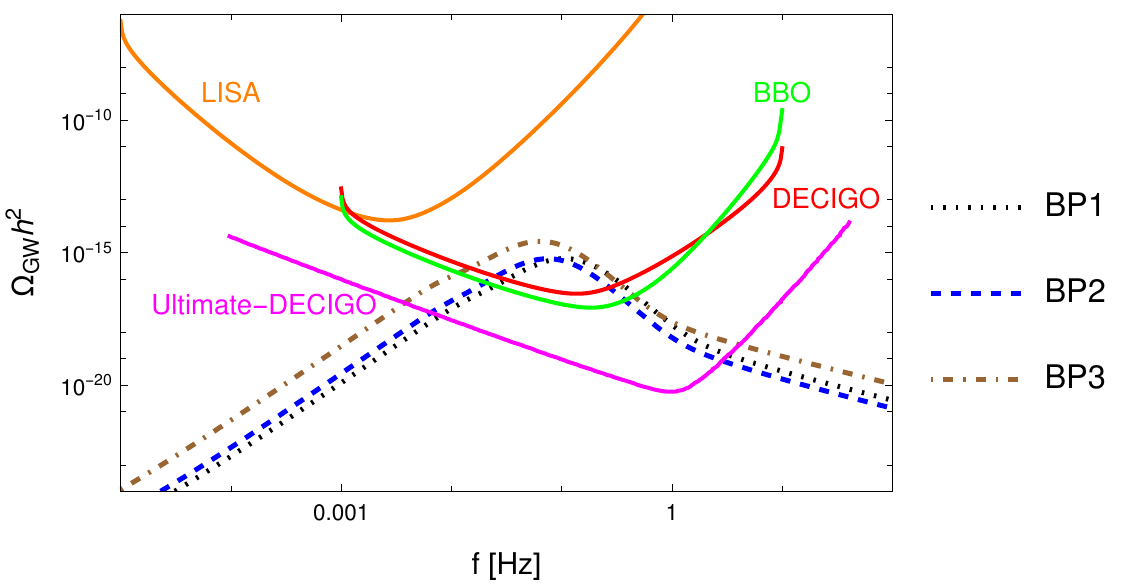}
    \caption{Left panel: Numerically computed $v_c/T_c$ values as a function of $\lambda_m \equiv \lambda_h-\lambda_{hD}^2/(4\lambda_D)$. Being in agreement with the analytical expression, strong FOPTs are achieved for small values of $\lambda_m$, or equivalently, small values of the dark $U(1)_D$ Higgs mass.  Right panel: FOPT-associated GW spectra for our three BPs summarised in Table~\ref{tab:BPs}.
The black dotted line corresponds to the first BP, the blue dashed line depicts the second BPs, and the brown dot-dashed line represents the third BP. The sensitivity curves of future space-base GW experiments, including LISA, BBO, DECIGO, and Ultimate-DECIGO, are shown as well}
    \label{fig:my_label}
\end{figure}
\section{Gravitational waves}
GWs produced by FOPTs have three main contributors: bubble wall collisions, sound waves in plasma, and magneto-hydrodynamic turbulence ~\cite{Kamionkowski:1993fg}. The sum of these three components can be computed using {\tt CosmoTransitions}. Fig. 3 shows the GW signals for three benchmark points (BPs) and the sensitivity of future space-based GW experiments such as LISA, DECIGO, and BBO. The three BPs account not only for neutrino masses and DM relic density, but also for strong FOPTs, and have different DM compositions (mostly WIMP/FIMP, or a similar contribution). All three BPs are within the reach of BBO, DECIGO, and Ultimate-DECIGO's detectability threshold.


\section{Conclusions}

This work discussed a model that extends the Standard Model to include dark matter and small neutrino masses using an extended seesaw framework. The model introduces two sets of three-generation neutrinos, with the third generation becoming FIMP-like particles. The heavier particle decays into the lighter one, making the lighter third-generation neutrino the FIMP dark matter candidate. The model also includes a WIMP dark matter candidate, the dark $U(1)_D$ gauge boson, creating a two-component WIMP-FIMP dark matter scenario. This study explored allowed parameter spaces and discusses prospects for detection in future experiments. It also showed that a first-order phase transition is possible in the scalar sector and that the model has the potential to generate stochastic gravitational waves that could be detected by future experiments. We have demonstrated that the strength of the electroweak first-order phase transition, quantified by the quantity $v_c/T_c$, where $T_c$ is the critical temperature and $v_c$ is the SM Higgs vacuum expectation value at $T_c$, may become larger than unity for small values of the dark $U(1)_D$ Higgs mass. Therefore, one of the essential ingredients for successful electroweak baryogenesis is achieved in our model. We presented benchmark points that demonstrate the model's potential detectability from GW observatories and DM experiments.
\begin{table}[t!]
\centering
\renewcommand{\arraystretch}{1.2}
\tabcolsep=0.1cm
\begin{tabular}{||c|c|c|c|c|c|c|c|c|c|c||}
\hline \hline
BPs & $v_D$ [TeV] & $M_{H_2}$ [GeV] &
$\sin \theta$ & 
$g_D$ [$10^{-4}$]  &
$\alpha$ & $\frac{\beta}{\mathcal{H}_*}$ & $T_n$ [GeV] &
$\frac{v_c}{T_c}$ &
$\frac{\Omega_{\rm WIMP}}{\Omega_{\rm Tot}}$ &
$\frac{\Omega_{\rm FIMP}}{\Omega_{\rm Tot}}$ 
\\ \hline
BP1 & 3.37 & 2.21 & 0.082 & 3.1 
& 0.238 & 13671 & 34.43 & 4.67 &
0.46 & 0.54 
\\ \hline
BP2 & 0.673& 2.77 & -0.076 & 19.7 
&0.139 & 6760.0 & 46.67 & 3.56 &
0.044 & 0.956
\\ \hline
BP3 & 4.63 & 1.0 & 0.060 & 1.0
& 0.461 & 13820 & 21.58 & 6.76 &
0.87 & 0.13\\
\hline \hline
\end{tabular}
\caption{Three BPs. The first four columns represent the input model parameters, the fifth, sixth, and seventh columns are GW-related quantities, and the eighth column shows the strength of the FOPT. The last two columns denote the WIMP and FIMP contributions to the total DM relic density $\Omega_{\rm Tot}h^2=0.12$; for the first BP, both the WIMP and FIMP equally contribute to the total DM relic density, while the second (third) BP is mostly composed of FIMP (WIMP) DM. In all three cases, the LFV bounds are satisfied, and the neutrino masses can successfully be generated. The GW signals corresponding to the three BPs are shown in Fig. 3.
}
\label{tab:BPs}
\end{table}

\section{Acknowledgments}
The results presented in this document have been derived in collaboration with J. Kim and S. Khan. This project has received funding from the European Unions Horizon 2020 research and innovation program under the Marie Skłodowska-Curie grant agreement No 860881-HIDDeN.

\bibliographystyle{JHEP}
\providecommand{\href}[2]{#2}\begingroup\raggedright\endgroup

\end{document}